\documentclass[conference]{IEEEtran}

\usepackage{amsmath}
\usepackage{graphicx}
\usepackage{subcaption}
\usepackage{algorithm}
\usepackage{algpseudocode}
\usepackage{cite}
\usepackage{array}
\usepackage{mathtools}
\usepackage[left=0.6in,right=0.6in,top=0.7in,bottom=0.7in]{geometry}

\newcolumntype{L}[1]{>{\raggedright\let\newline\\\arraybackslash\hspace{0pt}}m{#1}}
\newcolumntype{C}[1]{>{\centering\let\newline\\\arraybackslash\hspace{0pt}}m{#1}}
\newcolumntype{R}[1]{>{\raggedleft\let\newline\\\arraybackslash\hspace{0pt}}m{#1}}
\newcolumntype{P}[1]{>{\centering\arraybackslash}p{#1}}
\newcolumntype{M}[1]{>{\centering\arraybackslash}m{#1}}

\ifCLASSINFOpdf
\else
\fi
\begin{document}
%
\title{SNR Wall for Cooperative Spectrum Sensing Using Generalized Energy Detector}

\author{\IEEEauthorblockN{Kamal M. Captain}
	\IEEEauthorblockA{Dhirubhai Ambani Institute of \\ Information and communication Technology\\
		Gujarat, India 382007\\
		Email: kamalcaptain@daiict.ac.in}
	
	\and
	\IEEEauthorblockN{Manjunath V. Joshi}
	\IEEEauthorblockA{Dhirubhai Ambani Institute of \\ Information and communication Technology\\
		Gujarat, India 382007\\
		Email: mv\_joshi@daiict.ac.in}}


%


\maketitle

\begin{abstract}
Cognitive radio (CR) is a promising scheme to improve the spectrum utilization. Spectrum sensing (SS) is one of the main tasks of CR. Cooperative spectrum sensing (CSS) is used in CR to improve detection capability. Due to its simplicity and low complexity, sensing based on energy detection known as conventional energy detection (CED) is widely adopted. CED can be generalized by changing the squaring operation of the amplitude of received samples by an arbitrary positive power $p$ which is referred to as the generalized energy detector (GED). The performance of GED degrades when there exists noise uncertainty (NU). In this paper, we investigate the performance of CSS by considering the noise NU when all the secondary users (SUs) employ GED. We derive the signal to noise ratio (SNR) wall for CSS for both hard and soft decision combining. All the derived expressions are validated using Monte Carlo (MC) simulations. 
\end{abstract}


%
\IEEEpeerreviewmaketitle

\section{Introduction}
Cognitive radio (CR) has the potential to solve the spectrum scarcity problem by allowing the SUs to access the licensed band when primary users (PUs) are not using them \cite{mitola}. To access the unused licensed band, SU needs to check the occupancy status of the PU, which is termed in the literature as the spectrum sensing (SS) and is one of the main tasks of CR. 
In the literature different techniques for spectrum sensing have been investigated \cite{steven,cyclostationary,covariance1,eigenvalue1,urko,digham}. 
Energy detector \cite{urko,digham} is a popular SS technique since it does not require prior knowledge of the PU and is easy to implement. CED is generalized by replacing the squaring operation of the received signal amplitude by an arbitrary positive power $p$, which is referred to as the generalized energy detector (GED) \cite{sanket_ncc,sanket_apcc} or the  improved energy detector \cite{improved,improved_2} or $p$-norm detector \cite{p_norm,gaussian_approximation}. It is shown that performance of the energy detector can be improved by choosing a suitable value of $p$ \cite{improved,improved_1,improved_2}. 

In GED, the decision on the occupancy status of the PU channel is made based on a predefined threshold, which can be determined by the noise variance that plays an important role in determining the performance of the detector. One has to know the true noise variance to determine the value of this threshold. If this value is known exactly, one can sense the occupancy of a PU even at a very low SNR provided the sensing time is made sufficiently large \cite{steven}. However, in practice the noise variance varies with the time as well as the location and hence, it is difficult to find it's exact value. Due to this, there exists unpredictability about the true noise variance which is known as noise uncertainty (NU) because of which there exists a phenomenon called SNR wall \cite{SNR_wall}. It says that if the noise variance is not known exactly and is confined to an interval, one cannot achieve targeted detection performance when the SNR falls below certain value regardless of the sensing time. This makes CED an inefficient sensing method. Authors in \cite{SNR_wall} derive the SNR wall for CED. The effect of uniformly distributed NU is studied in \cite{reliability} and the expression for SNR wall is derived. In \cite{sanket_ncc,sanket_apcc}, the performance of GED is studied under uniformly distributed NU. It is shown in \cite{sanket_ncc} that under the worst case of NU the SNR wall is independent of $p$ and the CED represents the optimum energy detector. The expression for SNR wall is obtained in \cite{sanket_apcc} for the same scenario and it is shown that the SNR wall is independent of $p$.

The detection performance of the CSS under NU is studied in \cite{impact}. Authors in \cite{CSS_NU} propose CSS with adaptive thresholds to improve the detection performance. SNR wall for CSS with CED assuming the same SNR and NU for all the cooperating SUs (CSUs) is discussed in \cite{SNR_wall_CSS_1,SNR_wall_CSS_2}. However, in practice the SNR varies with the time and the location since it depends on the distance between the PU and the SU and the propagation path. Also, the NU depends on calibration error, variations in thermal noise and changes in low nose amplifier (LNA) gain. Hence, the assumption of the same SNR and NU at all the SUs is not valid in practice. The scenario in which different CSUs have the varying NU are studied in \cite{CSS_NU,impact} but they do not discuss the SNR wall. The discussion on SNR wall in \cite{SNR_wall_CSS_1} is limited to soft combining only whereas the same in \cite{SNR_wall_CSS_2} for hard combining is limited to AND combining rule only. Also in \cite{SNR_wall_CSS_1,SNR_wall_CSS_2}, all the CSUs use CED for detection. In this paper, we derive the expression for SNR wall when all the SUs use GED, without enforcing any assumption on the SNR and the uncertainty. We derive the SNR wall for hard as well as for soft combining. For hard combining we consider all three possible cases, i.e., OR, AND and $k$ out of $M$ combining rule. Note that, although authors in \cite{SNR_wall,reliability,sanket_ncc,sanket_apcc,SNR_wall_CSS_1,SNR_wall_CSS_2} discuss NU and SNR wall, their analysis is limited to real valued signal only. However, in practice SU receives complex valued signal. Hence, in this paper, we provide the analysis by considering the received signal as complex.
\section{Generalized Energy Detector Under Noise Uncertainty}
%
\subsection{System Model}
\label{system}
Let us consider that $M$ number of SUs are cooperating and each of them takes $N$ samples during the observation interval. Hence, the received signal at the $i^{\text{th}}$ SU can be written as
\begin{equation}
\label{system_model}
y_i(n)=
\begin{cases}
w_i(n); \ \ \ \ \ \ \ \ \ \ \ \ \ \ \ \ \ \ H_0,\\
h_i(n) s_i(n)+w_i(n); \ \ H_1,
\end{cases}
\end{equation} 
where $h_i(n), s_i(n)$ and $w_i(n)$ are the $n^{\text{th}}$ sample of the complex fading channel gain, PU signal and the noise, respectively, at the $i^{\text{th}}$ CSU with $n=1,2,\cdots,N$ and $i=1,2,\cdots,M$. The signal and the noise samples are independent and identically distributed (i.i.d) with $s_i(n)\sim \mathcal{C}\mathcal{N}(0,\,\sigma_{s_i}^{2})$\footnote{Complex Gaussian signal assumption is valid, for example, in an orthogonal frequency-division multiplexing signal having a large number of subcarriers \cite{complex_gaussian_1,complex_gaussian_2}, in frequency-shift keying signals that can be reasonably approximated as Gaussian process due to the complex time structure.} and $w_i(n)\sim \mathcal{C}\mathcal{N}(0,\,\sigma_{w_i}^{2})$. Here, the notation $\mathcal{C}\mathcal{N}(\bar{x},\,\sigma_x^{2})$ denotes complex Gaussian distribution with mean $\bar{x}$ and variance $\sigma_x^2$. In this paper we restrict our discussion to additive white Gaussian noise (AWGN) channel only and hence we consider $h_i(n)=1$. The hypotheses $H_0$ and $H_1$ correspond to free and occupied primary channel, respectively.

Now, considering that all the SUs employ GED, the decision statistic at the $i^{\text{th}}$ SU is given by
\begin{equation}
\label{TGED}
T_{i}=\frac{1}{N}\sum_{n=1}^N |y_i(n)|^p,
\end{equation}

\subsection{Noise Uncertainty Model}
The characterization of AWGN, i.e., $w_i(n)$, in Eq. (\ref{system_model}) depends on its variance. In general, when we consider different detection methods, it is assumed that the true noise variance at the input of SU is known a priori. The same is used in choosing a threshold for detecting the presence or the absence of a PU signal. However, in practice, the noise variance varies over time and location resulting in NU \cite{SNR_wall,reliability}. 

The average or the expected value of the noise variance $\hat{\sigma}_{w_i}^2$ is known at the $i^{\text{th}}$ SU. Let $\sigma_{w_i}^2$ be the true noise variance at the $i^{\text{th}}$ SU which may vary from $\hat{\sigma}_{w_i}^2$ giving rise to noise uncertainty (NU). The NU factor $\beta_i$ at the $i^{\text{th}}$ SU is defined as $\beta_i=\frac{\hat{\sigma}_{w_i}^2}{\sigma_{w_i}^2}$ which is a random variable since $\sigma_{w_i}^2$ is random. Let the upper bound on the NU be $L_i$ dB, which can be written as $L_i=sup\left\{10log_{10}\beta_i \right\}$. 
Assuming that the $\beta_i$ in dB is uniformly distributed in the range $[-L_i,L_i]$ \cite{SNR_wall}, implying that it is restricted in the range $[10^{\frac{-L_i}{10}},10^{\frac{L_i}{10}}]$. The probability density function (pdf) of $\beta_i$ can be written as
\begin{equation}
	\label{pdf_B_i}
	f_{\beta_i}(x)=
	\begin{cases}
		0, \ \ \ \ \ \ \ \ \ \ \ \ \ \ x<10^{\frac{-L_i}{10}} \\
		\frac{5}{\left[ln(10)\right]L_ix}, \ \ \ 10^{\frac{-L_i}{10}}<x<10^{\frac{L_i}{10}} \\
		0, \ \ \ \ \ \ \ \ \ \ \ \ \ \ x>10^{\frac{L_i}{10}}
	\end{cases}
\end{equation} 
where, $ln(z)$ represents the natural logarithm of $z$.
\subsection{Detection Probabilities}
\label{detection_probabilities}
When there is no cooperative sensing, the performance of the $i^{\text{th}}$ SU is measured in terms of probability of false alarm $(P_{F_i})$ and the probability of detection $(P_{D_i})$ which are defined as $P_{F_i}=Pr\left\{T_{i}>\tau | H_0 \right\}$ and $P_{D_i}=Pr\left\{T_{i}>\tau | H_1 \right\}$, respectively, 
where $\tau$ and $Pr\left\{\cdot\right\}$ represent the decision threshold and the probability operator, respectively.

If $N$ is chosen relatively large then by using central limit theorem (CLT), the pdf of the decision statistic given in Eq. (\ref{TGED}) can be modeled by Gaussian distribution \cite{sanket_ncc,sanket_apcc,p_norm,SNR_wall_CSS_1,SNR_wall_CSS_2}. In this case the pdf can be represented by mean and variance only. Therefore considering uncertainty factor $\beta_i$, the mean and variance at $i^{\text{th}}$ SU can be given as
\begin{equation}
	\label{mean_var_H0}
	\mu_{0_i}=G_p\sigma_{w_i}^p, \ \ \sigma_{0_i}^2=\frac{K_p}{N}\sigma_{w_i}^{2p},
\end{equation} 
\begin{equation}
	\label{mean_var_H1}
	\mu_{1_i}=G_p(1+\beta_i\gamma_i)^{\frac{p}{2}}\sigma_{w_i}^p, \ \ \sigma_{1_i}^2=\frac{K_p}{N}(1+\beta_i\gamma_i)^p\sigma_{w_i}^{2p},
\end{equation} 
where $\mu_{0_i}$, $\sigma_{0_i}^2$ and $\mu_{1_i}$, $\sigma_{1_i}^2$ correspond to the mean and the variance under $H_0$ and $H_1$, respectively. Here, $\gamma_i$ is the average SNR at the $i^{\text{th}}$ SU, $G_p=\Gamma\left(\frac{p+2}{2}\right)$, $K_p=\Gamma(p+1)-\Gamma^2\left(\frac{p+2}{2}\right)$, where $\Gamma(a)$ represents the complete Gamma function \cite[6.1.1]{handbook}. Using these, $P_{F_i}$ and $P_{D_i}$ for the $i^{\text{th}}$ CSU when we consider a fixed value of NU factor can be obtained as
\begin{equation}
	\label{PFi}
	P_{F_i}=Q\Bigg(\frac{\tau-\mu_{0_i}}{\sigma_{0_i}}\Bigg), \ \text{and} \ P_{D_i}=Q\Bigg(\frac{\tau-\mu_{1_i}}{\sigma_{1_i}}\Bigg),
\end{equation}
where $Q(t)=\frac{1}{\sqrt{2\pi}}\int_t^\infty e^{-(\frac{x^2}{2})}dx$.

The threshold $\tau$ is chosen as $\lambda \hat{\sigma}_w^p$ for GED, where $\lambda> 0$ is a constant. 
We assume that $\hat{\sigma}_{w_1}^2=\hat{\sigma}_{w_2}^2= \cdots=\hat{\sigma}_{w_M}^2$ and hence $\tau$ is same for all the CSUs. 
$\beta_i$ being a random variable, one can obtain the average $P_{F_i}$ and $P_{D_i}$, i.e., $\bar{P}_{F_i}$ and $\bar{P}_{D_i}$, by using the means and variances from Eq. (\ref{mean_var_H0}) and Eq. (\ref{mean_var_H1}) in Eq. (\ref{PFi}) and averaging them over the pdf of $\beta_i$ given in Eq. (\ref{pdf_B_i}). Therefore, $\bar{P}_{F_i}$ and $\bar{P}_{D_i}$ for the $i^{\text{th}}$ CSU can be obtained as
\begin{equation}
	\small
	\label{bar_PFi}
	\bar{P}_{F_i}=\int_a^b Q\left(\left(\lambda x^{\frac{p}{2}}-G_p\right)\sqrt{\frac{N}{K_p}}\right)\frac{5}{L_ixln(10)}dx, \ \text{and}
\end{equation}
\begin{equation}
	\small
	\label{bar_PDi}
	\bar{P}_{D_i}=\int_a^b Q\left(\frac{\lambda x^{\frac{p}{2}}-G_p(1+x\gamma_i)^{\frac{p}{2}}}{(1+x\gamma_i)^{\frac{p}{2}}}\sqrt{\frac{N}{K_p}}\right)\frac{5}{L_ixln(10)}dx,
\end{equation}
where, $a=10^{\frac{-L_i}{10}}$ and $b=10^{\frac{L_i}{10}}$. Note that, the integrals in Eq. (\ref{bar_PFi}) and Eq. (\ref{bar_PDi}) can be reduced to closed form using an approximation to $Q(\cdot)$ function. The goal of this paper is to derive the SNR wall and to do that the equations in integral are sufficient. Hence, we keep these equations in integral form only. 
\section{SNR Wall for Cooperative Spectrum Sensing}
\label{SNR_wall}
When we consider CSS, we get a combined average probability of false alarm ($\bar{Q}_F$) and average probability of detection ($\bar{Q}_D$) in each case based on the number of CSUs. Given different SNRs $(\gamma_i>0)$ at the SUs, $ i=1,2,\cdots,M$, if there exists a threshold for which 
\begin{equation}
	\label{condition}
	\lim \limits_{N\rightarrow \infty}\bar{Q}_{F}=0 \ \text{and} \ \lim \limits_{N\rightarrow \infty}\bar{Q}_{D}=1,
\end{equation}
then the sensing scheme is considered as unlimitedly reliable \cite{reliability}. In other words, if the channel is sensed for sufficiently long time, i.e., $N\rightarrow \infty$, one can achieve the desired target $\bar{Q}_F=0$ and $\bar{Q}_D=1$ at any SNR level. However, this is possible only when there is no NU. In the presence of NU, it is not possible to achieve unlimited reliability below certain SNR value even when $N$ is very large, i.e., $N\rightarrow \infty$ \cite{reliability}. The SNR value below which it is not possible to achieve an unlimited reliability is referred as the SNR wall \cite{reliability} and in this case at least one of the conditions in Eq. (\ref{condition}) is not satisfied. However, when the SNR is above the SNR wall, there exists a threshold $\tau$ for which both the conditions in Eq. (\ref{condition}) are satisfied.

In this section, we derive the SNR wall for CSS under NU by considering hard as well as soft decision combining.
\subsection{Hard Decision Combining}
In hard decision combining all the CSUs take decisions on the occupancy of the channel and send their results as ON/OFF to the fusion center (FC). The FC then takes the final decision considering all the received decisions. In this case, we investigate the SNR wall for three combining rules, i.e., OR, AND and $k$ out of $M$ combining rule. 
\subsubsection{OR Combining Rule}
\label{OR_rule}
In OR combining rule, the FC declares the PU as active whenever at least one of the CSUs reports the channel as occupied. Considering this, we first derive the SNR wall for $M=2$ only and then extend the result to any number of CSUs. Let $L_1$ and $L_2$ be the upper bounds on the NU factors and $\gamma_1$ and $\gamma_2$ be the SNRs at the two CSUs. In this case with $M=2$, $\bar{Q}_F$ and $\bar{Q}_D$ at the FC can be written as
\begin{equation}
\small
\label{Q_F_OR}
\bar{Q}_F=\bar{P}_{F_1}+\bar{P}_{F_2}-\bar{P}_{F_1}\bar{P}_{F_2} \ \text{and} \ \bar{Q}_D=\bar{P}_{D_1}+\bar{P}_{D_2}-\bar{P}_{D_1}\bar{P}_{D_2}.
\end{equation}
To derive the SNR wall, we make use of the following result,
\begin{equation}
\label{limit}
\lim \limits_{N\rightarrow \infty} Q\left(a\sqrt{N}\right)=
\begin{cases}
0, \ \ \ \text{if} \ a>0, \\
1, \ \ \ \text{if} \ a<0, \\
0.5 \ \ \text{if} \ a=0.
\end{cases}
\end{equation} 
Since $\hat{\sigma}_{w_i}^2$s are known, we first need to find $\lambda$ for which the conditions in Eq. (\ref{condition}) are satisfied. 
From Eq. (\ref{Q_F_OR}), it is clear that to satisfy $\lim\limits_{N\rightarrow \infty} \bar{Q}_{F}=0$, we need both $\bar{P}_{F_1}$ and $\bar{P}_{F_2}$ to be 0. Hence, using Eq. (\ref{bar_PFi}) and Eq. (\ref{limit}), one has to set the $\lambda$ at both the CSUs as
\begin{equation}
\label{OR_4}
\lambda\geq G_p\left(10^\frac{L_1}{10}\right)^{\frac{p}{2}} \ \text{AND} \ \lambda\geq G_p\left(10^\frac{L_2}{10}\right)^{\frac{p}{2}}.
\end{equation}
The condition in Eq. (\ref{OR_4}) can be written in compact form as
\begin{equation}
\label{OR_1}
\lambda \geq max \left\{G_p\left(10^{\frac{L_1}{10}}\right)^{\frac{p}{2}},G_p\left(10^{\frac{L_2}{10}}\right)^{\frac{p}{2}}\right\}
\end{equation}
Similarly, to satisfy the condition $\lim\limits_{N\rightarrow \infty} \bar{Q}_{D}=1$, we see from Eq. (\ref{Q_F_OR}) that $P_{D_1}$ or $P_{D_2}$ must be 1. Once again, using the Eq. (\ref{bar_PDi}) and the Eq. (\ref{limit}), we need to set $\lambda$ as
\begin{equation}
\small
\label{OR_2}
\lambda \leq G_p\left(10^{\frac{-L_1}{10}}+\gamma_1\right)^{\frac{p}{2}} \ \text{OR} \ \lambda \leq G_p\left(10^{\frac{-L_2}{10}}+\gamma_2\right)^{\frac{p}{2}}.
\end{equation}
If we assume $L_1>L_2$, then using the Eq. (\ref{OR_1}) and Eq. (\ref{OR_2}), $\lambda$ to be chosen for unlimited reliability should satisfy
\begin{equation}
\label{OR_condition_1}
\begin{aligned}
&G_p\left(10^{\frac{L_1}{10}}\right)^{\frac{p}{2}}\leq \lambda \leq G_p\left(10^{\frac{-L_1}{10}}+\gamma_1\right)^{\frac{p}{2}}, \ \text{OR} \\
&G_p\left(10^{\frac{L_1}{10}}\right)^{\frac{p}{2}}\leq \lambda \leq G_p\left(10^{\frac{-L_2}{10}}+\gamma_2\right)^{\frac{p}{2}}.
\end{aligned}
\end{equation}
Using Eq. (\ref{OR_condition_1}), the condition on $\gamma_1$ and $\gamma_2$ can be given as
\begin{equation}
\label{OR_SNR_Wall}
\gamma_1\geq 10^{\frac{L_1}{10}}-10^{\frac{-L_1}{10}} \ \ \text{OR} \ \ \gamma_2\geq 10^{\frac{L_1}{2}}-10^{\frac{-L_2}{10}}.
\end{equation}
Therefore the SNR wall for the OR case is obtained by considering equality condition in Eq. (\ref{OR_SNR_Wall}).  

To understand this, let us take $L_1=1 \ dB$ and $L_2=0.5 \ dB$. Substituting in Eq. (\ref{OR_SNR_Wall}), we get $\gamma_1=0.4646$ and $\gamma_2=0.3676$. Therefore one can achieve unlimited reliability if $\gamma_1 \geq 0.4646$ or $\gamma_2 \geq 0.3676$. One can also see from Eq. (\ref{OR_SNR_Wall}) that the SNR wall in this case is independent of $p$. 

Following a similar procedure, the conditions given for the case of $M=2$ in Eq. (\ref{OR_SNR_Wall}) can be extended to any $M$ as
\begin{equation}
\label{OR_wall_M}
\gamma_i\geq 10^{\frac{L^+}{10}}-10^{\frac{-L_i}{10}}, \ \text{for} \ i=1,2,\cdots,M,
\end{equation}
where $L^+=\max\left\{L_1,L_2,\cdots,L_M\right\}$. In this case, to achieve unlimited reliability, any one among $M$ conditions in Eq. (\ref{OR_wall_M}) must be satisfied. 

The scenario when all the CSUs experience the same SNRs, i.e., $\gamma_1=\gamma_2=\gamma$, is discussed in \cite{SNR_wall_CSS_1,SNR_wall_CSS_2}. Considering this, the condition given in Eq. (\ref{OR_1}) remains the same since it does not involve $\gamma$ while that given in Eq. (\ref{OR_2}) can be rewritten as
\begin{equation}
\label{OR_3}
\gamma \leq \max \left\{G_p\left(10^{\frac{-L_1}{10}}+\gamma\right)^{\frac{p}{2}},G_p\left(10^{\frac{-L_2}{10}}+\gamma\right)^{\frac{p}{2}}\right\}.
\end{equation} 
Once again, assuming $L_1>L_2$ and using Eq. (\ref{OR_1}) and Eq. (\ref{OR_3}), the SNR wall can be obtained as
\begin{equation}
\gamma = 10^{\frac{L_1}{10}}-10^{\frac{-L_2}{10}}.
\end{equation}
Following a similar procedure, the SNR wall for $M$ CSUs with $\gamma_1=\gamma_2= \cdots = \gamma_M$ can be obtained as
\begin{equation}
\label{OR_equal_snr}
\gamma = 10^{\frac{L^+}{10}}-10^{\frac{-L^-}{10}},
\end{equation}
where $L^-=\min\left\{L_1,L_2,\cdots,L_M\right\}$.
\subsubsection{AND Rule}
Here, the FC declares the channel as occupied only when all the CSUs PU channel as occupied. Similar to OR case, here also we first derive SNR wall by considering $M=2$ and then extend it to any $M$. With $M=2$, $\bar{Q}_F$ and $\bar{Q}_D$ can be written as
\begin{equation}
\label{Q_F_AND}
\bar{Q}_F=\bar{P}_{F_1}\bar{P}_{F_2} \ \text{and} \  \bar{Q}_D=\bar{P}_{D_1}\bar{P}_{D_2}.
\end{equation}
It is clear from Eq. (\ref{Q_F_AND}) that in order to satisfy the condition on $\bar{Q}_F$ in Eq. (\ref{condition}), either $\bar{P}_{F_1}$ or $\bar{P}_{F_2}$ must be 0. Hence, one has to select $\lambda$ as
\begin{equation}
\label{AND_1}
\lambda \geq \min \left\{G_p\left(10^{\frac{L_1}{10}}\right)^{\frac{p}{2}},G_p\left(10^{\frac{L_2}{10}}\right)^{\frac{p}{2}}\right\}.
\end{equation}
Similarly, to satisfy the condition on $\bar{Q}_D$, both $\bar{P}_{D_1}$ and $P_{\bar{D}_2}$ in Eq. (\ref{Q_F_AND}) must be 1 and hence we need to set $\lambda$ as
\begin{equation}
\label{AND_2}
\lambda \leq G_p\left(10^{\frac{-L_1}{10}}+\gamma_1\right)^{\frac{p}{2}} \ \text{AND} \ \lambda \leq G_p\left(10^{\frac{-L_2}{10}}+\gamma_2\right)^{\frac{p}{2}}.
\end{equation}
Once again assuming $L_1>L_2$ and using Eq. (\ref{AND_1}) and Eq. (\ref{AND_2}),  $\lambda$ has to be selected as
\begin{equation}
\begin{aligned}
&G_p\left(10^{\frac{L_2}{10}}\right)^{\frac{p}{2}} \leq \lambda \leq G_p\left(10^{\frac{-L_1}{10}}+\gamma_1\right)^{\frac{p}{2}}, \ \text{AND} \\
&G_p\left(10^{\frac{L_2}{10}}\right)^{\frac{p}{2}} \leq \lambda \leq G_p\left(10^{\frac{-L_2}{10}}+\gamma_2\right)^{\frac{p}{2}}.
\end{aligned}
\end{equation}
Using this, $\gamma_1$ and $\gamma_2$ in this case should satisfy
\begin{equation}
\label{SNR_wall_AND}
\gamma_1\geq 10^{\frac{L_2}{10}}-10^{\frac{-L_1}{10}} \ \ \text{AND} \ \ \gamma_2\geq 10^{\frac{L_2}{10}}-10^{\frac{-L_2}{10}}.
\end{equation}
From this, the equality condition in the Eq. (\ref{SNR_wall_AND}) gives us the SNR walls for the two CSUs. Once again considering $L_1=1 \ dB$ and $L_2=0.5 \ dB$, the unlimitedly reliable performance can be obtained if $\gamma_1\geq 0.3277$ and $\gamma_2\geq 0.2308$. Note that, in this case both the SNRs have to satisfy the inequality conditions. Once again, the conditions given in Eq. (\ref{SNR_wall_AND}) can be extended to any number of $M$ and is given by Eq. (\ref{OR_wall_M}) with $L^+=\min\left\{L_1,L_2,\cdots,L_M\right\}$.
Note that all the SNRs must be $\geq$ their respective SNR walls in order to achieve unlimited reliability. 

When $\gamma_1=\gamma_2=\gamma$, the Eq. (\ref{AND_1}) remains the same but the Eq. (\ref{AND_2}) can be rewritten as
\begin{equation}
\label{AND_3}
\lambda \leq \min \left\{G_p\left(10^{\frac{-L_1}{10}}+\gamma\right)^{\frac{p}{2}},G_p\left(10^{\frac{-L_2}{10}}+\gamma\right)^{\frac{p}{2}}\right\}
\end{equation}
With $L_1>L_2$ and using Eq. (\ref{AND_1}) and Eq. (\ref{AND_3}), the SNR wall in this case can be obtained as
\begin{equation}
\gamma = 10^{\frac{L_2}{10}}-10^{\frac{-L_1}{10}}.
\end{equation}
Considering $M$ CSUs with $\gamma_1=\gamma_2,\cdots,\gamma_M=\gamma$, the SNR wall can be given by Eq. (\ref{OR_equal_snr}) with $L^+=\min\left\{L_1,L_2,\cdots,L_M\right\}$ and $L^-=\max\left\{L_1,L_2,\cdots,L_M\right\}$.
\subsubsection{$k$ Out Of $M$ Combining Rule}
In this rule, FC declares the channel as occupied when $k$ out of the total of $M$ CSUs report the PU channel as occupied. For this case, we first derive the SNR wall by considering $M=3$ and $k=2$, and then extend the result to general case of any $M$ and $k$. With this setting, $\bar{Q}_F$ and $\bar{Q}_D$ can be written as
\begin{equation}
\label{Q_F_k_out_M}
\bar{Q}_F=\bar{P}_{F_1}\bar{P}_{F_2}+\bar{P}_{F_2}\bar{P}_{F_3}+\bar{P}_{F_1}\bar{P}_{F_3}-2\bar{P}_{F_1}\bar{P}_{F_2}\bar{P}_{F_3},
\end{equation}
\begin{equation}
\label{Q_D_k_out_M}
\bar{Q}_D=\bar{P}_{D_1}\bar{P}_{D_2}+\bar{P}_{D_2}\bar{P}_{D_3}+\bar{P}_{D_1}\bar{P}_{D_3}-2\bar{P}_{D_1}\bar{P}_{D_2}\bar{P}_{D_3},
\end{equation}
Now for $\lim\limits_{N\rightarrow \infty} \bar{Q}_{F}=0$, we must have any of the two $\bar{P}_{F_i}\text{s}, \ i=1,2,3$ must be $0$ in Eq. (\ref{Q_F_k_out_M}). 
Therefore, $\lambda$ has to be selected such that
\begin{equation}
\label{k_out_M_7}
\begin{aligned}
\lambda &\geq \max \left\{G_p\left(10^{\frac{L_1}{10}}\right),G_p\left(10^{\frac{L_2}{10}}\right)\right\}, \ \text{OR} \\
\lambda &\geq \max \left\{G_p\left(10^{\frac{L_2}{10}}\right),G_p\left(10^{\frac{L_3}{10}}\right)\right\}, \ \text{OR} \\
\lambda &\geq \max \left\{G_p\left(10^{\frac{L_1}{10}}\right),G_p\left(10^{\frac{L_3}{10}}\right)\right\}.
\end{aligned}
\end{equation} 
To achieve the other condition of $\lim\limits_{N\rightarrow \infty} \bar{Q}_{D}=1$, using Eq. (\ref{Q_D_k_out_M}), any two $\bar{P}_{D_i}\text{s}$, for $i=1,2,3$ must be $1$ which is obtained by setting $\lambda$ as 
To achieve this, $\lambda$ has to be selected as
\begin{equation}
\label{k_out_M_8}
\begin{aligned}
\lambda &\leq \min\left\{G_p\left(10^{\frac{-L_1}{10}}+\gamma_1\right)^{\frac{p}{2}},G_p\left(10^{\frac{-L_2}{10}}+\gamma_2\right)^{\frac{p}{2}}\right\} \ \text{OR} \\
\lambda &\leq \min\left\{G_p\left(10^{\frac{-L_1}{10}}+\gamma_1\right)^{\frac{p}{2}},G_p\left(10^{\frac{-L_3}{10}}+\gamma_3\right)^{\frac{p}{2}}\right\} \ \text{OR} \\
\lambda &\leq \min\left\{G_p\left(10^{\frac{-L_2}{10}}+\gamma_2\right)^{\frac{p}{2}},G_p\left(10^{\frac{-L_3}{10}}+\gamma_3\right)^{\frac{p}{2}}\right\}
\end{aligned}
\end{equation}
In order to see the implications of these conditions, let us consider $L_1>L_2>L_3$. Using Eq. (\ref{k_out_M_7}) and Eq. (\ref{k_out_M_8}), the conditions on $\gamma_1$, $\gamma_2$ and $\gamma_3$ can be given by
\begin{equation}
\small
\label{SNR_Wall_K_out_M}
\gamma_1\geq10^{\frac{L_2}{10}}-10^{\frac{-L_1}{10}}, \gamma_2\geq10^{\frac{L_2}{10}}-10^{\frac{-L_2}{10}},  \gamma_3\geq10^{\frac{L_2}{10}}-10^{\frac{-L_3}{10}}.
\end{equation}
Therefore, for $k=2$ any two conditions given in Eq. (\ref{SNR_Wall_K_out_M}) must be satisfied, in order to get unlimited reliability. Equality sign in Eq. (\ref{SNR_Wall_K_out_M}) then gives us the SNR wall for $2$ out of $3$ rule. One can also see from Eq. (\ref{SNR_Wall_K_out_M}) that the SNR wall in this case is independent of the value of $p$. As an example, let us take $L_1=1 \ dB$, $L_2=0.7 \ dB$ and $L_3=0.5 \ dB$. Substituting in Eq. (\ref{SNR_Wall_K_out_M}), we get the SNR walls for $3$ CSUs as $\gamma_1=0.3806$, $\gamma_2=0.3238$ and $\gamma_3=0.2836$. Therefore one can achieve unlimited reliability if any two of the SNRs at the CSUs are $\geq$ to their respective SNR wall values. 

Following the similar procedure, the conditions given for the case of $M=3$ in Eq. (\ref{SNR_Wall_K_out_M}) can be extended to any $k$ out of $M$ CSUs and is given by Eq. (\ref{OR_wall_M}) with 
$L^+=\min\left\{k \ \text{largest from} \ (L_1,L_2,\cdots,L_M)\right\}$. 
For example, with $M=3$, $k=2$ and $L_1>L_2>L_3$ then $L^+=L_2$ and we arrive at Eq. (\ref{SNR_Wall_K_out_M}). Note that, to achieve unlimited reliability, any $k$ SNRs must be $\geq$ their respective SNR walls. 

The equal SNR wall scenario when $\gamma_1=\gamma_2=\gamma_3=\gamma$ can be derived using Eq. (\ref{k_out_M_7}) and Eq. (\ref{k_out_M_8}). Assuming $L_1>L_2>L_3$, the SNR wall for $M=3$ and $k=2$ can be obtained as
\begin{equation}
\gamma = 10^{\frac{L_2}{10}}-10^{\frac{-L_2}{10}}
\end{equation}
Now considering $\gamma_1=\gamma_2=\cdots=\gamma_M$, the SNR wall for the general case can be given by Eq. (\ref{OR_equal_snr}) with $L^+=\min\left\{L_1,L_2,\cdots,L_M\right\}$ and $L^-=\max\left\{k \ \text{smallest from} \ (L_1,L_2,\cdots,L_M)\right\}$.
For example, with $M=3$, $k=2$ and $L_1>L_2>L_3$ we have $L^+=L^-=L_2$. Note that, the SNR wall for OR and AND combining can be obtained as the special cases of $k$ out of $M$ combining rule. Choosing $k=1$ and $k=M$ result in OR and AND combining rules, respectively.

\begin{table}
	\renewcommand{\arraystretch}{1.2}
	\caption{Comparison of SNR walls for hard combining. Here, $M=3$, $L_1=1 \ dB$, $L_2=0.7 \ dB$ and $L_3=0.5 \ dB$.}
	\begin{tabular*}\linewidth{ |C{0.15\linewidth} | C{0.15\linewidth} |C{0.15\linewidth} |C{0.15\linewidth}|C{0.15\linewidth}| }
		\hline
		Decision Rule & $\gamma_1$ &$\gamma_2$& $\gamma_3$ & $k$ \\
		\hline
		OR&   $0.4646$  & $0.4077$   &$0.3677$ & $1$ \\
		\hline
		AND &$0.3277$  & $0.2708$&  $0.2307$ & $3$\\
		\hline
		$2$ out of $3$ & $0.3806$    &$0.3238$&  $0.2836$ & $2$ \\
		\hline
	\end{tabular*}
\end{table}
In TABLE I, we list the SNR wall under OR, AND  and $k$ out of $M$ combining rule when hard combining is used. We consider $k=2$ for $k$ out of $M$ combining rule. Note that, the value of $k$ also represents the required number of SNRs are to be $\geq$ their respective SNR walls at the CSUs in order to get the unlimited reliability. Looking at Table I, one may notice that, though the SNR wall values that we get for OR combining rule are higher when compared to other two rules, it requires only a one SNR to be $\geq$ the respective SNR wall value to achieve unlimited reliability. When AND combining rule is used, the SNR wall values are smallest but we require all three SNRs $\geq$ their SNR wall values for achieving unlimited reliability. With $k$ out of $M$ combining rule, the SNR wall values lie between those of OR and AND combining rules, and any $k$ SNR values at the CSUs have to be $\geq$ their respective SNR wall values.
\subsection{Soft Decision Combining}
\label{soft_derivation}
We investigate the SNR wall for soft decision combining when equal gain combining (EGC) is used at the FC. Here, the decision on PU being ON/OFF is not taken by the CSUs. Instead, the decision statistic from all the CSUs are sent to the FC where they are added to obtain a new decision statistic and the decision is taken by FC based this. Let $T_i$ be the decision statistic at the $i^{\text{th}}$ CSU. Then, the new decision statistic at the FC is obtained as
\begin{equation}
\label{Statistic_SC}
T=\frac{1}{M}\sum_{i=1}^M T_i.
\end{equation}
To make it simple, we first carry out the derivations for $\bar{Q}_f$ and $\bar{Q}_d$ using two CSUs only, i.e., $M=2$ and then extend it to any $M$. We know that the decision statistics $T_i$ at two CSUs with $i=1,2$, respectively, follow Gaussian distribution with the mean and variance as given in Eq. (\ref{mean_var_H0}) and Eq. (\ref{mean_var_H1}). We can compute $T$ using Eq. (\ref{Statistic_SC}). $T$ is also Gaussian with mean and variance as
\begin{equation}
\small
\label{mu_0_css}
\mu_{0,c}=\frac{G_p}{2}\left[\sigma_{w_1}^p+\sigma_{w_2}^p\right], \ \text{and} \  \sigma_{0,c}^2=\frac{G_pK_p}{2^2N}\left[\sigma_{w_1}^{2p}+\sigma_{w_2}^{2p}\right],
\end{equation}
under $H_0$ and
\begin{equation}
\label{mu_sigma_1_css}
\begin{aligned}
\mu_{1,c}=\frac{G_p}{2}\left[(1+\beta_1\gamma_1)^{\frac{p}{2}}\sigma_{w_1}^p+(1+\beta_2\gamma_2)^{\frac{p}{2}}\sigma_{w_2}^p\right], \\
\sigma_{1,c}^2=\frac{G_pK_p}{2^2N}\left[(1+\beta_1\gamma_1)^p\sigma_{w_1}^{2p}+(1+\beta_2\gamma_2)^p\sigma_{w_2}^{2p}\right],
\end{aligned}
\end{equation}
under $H_1$, respectively. Here, subscripts $0,c$ and $1,c$ represent that the means and variances are for CSS under $H_0$ and $H_1$, respectively.

Using this, the probability of false alarm $(Q_F)$ and the probability of detection $(Q_D)$ for fixed values of $\beta_1$ and $\beta_2$ can be obtained using Eq. (\ref{PFi}). We know that with no cooperation, the threshold $\tau$ is chosen as $\lambda \hat{\sigma}_w^p$. Hence, when there are two CSUs, the threshold should be selected as $\frac{\lambda}{2}\left(\hat{\sigma}_{w_1}^p+\hat{\sigma}_{w_2}^p\right)$. Using this $\tau$ in Eq. (\ref{PFi}) and $\mu_{0,c}$ and $\sigma_{0,c}^2$ from Eq. (\ref{mu_0_css}), one can obtain $Q_F$ for fixed $\beta_1$ and $\beta_2$ after few manipulations as
\begin{equation}
\label{Q_F_fixed}
Q_F=Q\Bigg(\frac{2\lambda\beta_1^{\frac{p}{2}}\beta_2^{\frac{p}{2}}-G_p\left(\beta_1^{\frac{p}{2}}+\beta_2^{\frac{p}{2}}\right)}{\sqrt{\beta_1^p+\beta_2^p}}\sqrt{\frac{N}{K_p}}\Bigg),
\end{equation}
where, we used $\hat{\sigma}_{w_1}^2=\hat{\sigma}_{w_2}^2=1$\footnote{In literature when the NU is not considered, the expected value of noise variance is considered as the true noise variance. Here, the expected value of variances are assumed to be 1 for mathematical simplicity. In \cite{digham}, the noise variance is assumed to be 1.} and $\beta_1=\frac{\hat{\sigma}_{w_1}^2}{\sigma_{w_1}^2}$ and $\beta_2=\frac{\hat{\sigma}_{w_2}^2}{\sigma_{w_2}^2}$. Similarly, $Q_D$ for fixed $\beta_1$ and $\beta_2$ can be obtained by making use of mean and variance from Eq. (\ref{mu_sigma_1_css}) in Eq. (\ref{PFi}) as
\begin{equation}
\label{Q_D_fixed}
Q_D=Q\left(\frac{2\lambda\beta_1^{\frac{p}{2}}\beta_2^{\frac{p}{2}}-\frac{G_p\left(1+\beta_1\gamma_1\right)^{\frac{p}{2}}}{\beta_2^{-\frac{p}{2}}}-\frac{G_p\left(1+\beta_2\gamma_2\right)^{\frac{p}{2}}}{\beta_1^{-\frac{p}{2}}}}{\sqrt{\frac{K_p}{N}}\sqrt{(1+\beta_1\gamma_1)^p\beta_2^p+(1+\beta_2\gamma_2)^p\beta_1^p}}\right).
\end{equation}
$\beta$ being a random variable, one can obtain $\bar{Q}_{F}$ and $\bar{Q}_D$ by averaging $Q_{F}$ and $Q_{D}$ over joint pdf of $\beta_1$ and $\beta_2$. Assuming that $\beta_1$ and $\beta_2$ are independent and using Eq. (\ref{pdf_B_i}) the joint pdf of $\beta_1$ and $\beta_2$ is given by
\begin{equation}
\label{jpdf}
f(x,y)=
\begin{cases}
\ \ \ \ \ \ 0, \ \ \ \ \ \ \ \ \ \ \ \ \ x<a_1, y<a_2,  \\
\frac{25}{\left[ln(10)\right]^{2}L_1L_2xy}, \ \ a_1<x<b_1, a_2<y<b_2 \\
\ \ \ \ \ \ 0, \ \ \ \ \ \ \ \ \ \ \ \ \ x>b_1, y>b_2,
\end{cases}
\end{equation}
where, $L_1$ and $L_2$ correspond to the upper bound on $\beta$ in dB at CSUs 1 and 2, respectively. Here, $a_1=10^\frac{-L_1}{10}$, $b_1=10^\frac{L_1}{10}$, $a_2=10^\frac{-L_2}{10}$ and $b_2=10^\frac{L_2}{10}$. Using this, $\bar{Q}_{F}$ is obtained as
\begin{equation}
\small
\label{Q_F_average}
\bar{Q}_{F}=\int\limits_{a_1}^{b_1}\int\limits_{a_2}^{b_2} Q\left(\frac{\lambda\left(2(xy)^{\frac{p}{2}}\right)-G_p\left(x^{\frac{p}{2}}+y^{\frac{p}{2}}\right)}{\sqrt{\frac{K_p}{N}}\sqrt{x^p+y^p}}\right) f(x,y) dy dx.
\end{equation}
Similarly, one can obtain $\bar{Q}_{D}$ by averaging $Q_D$ in Eq. (\ref{Q_D_fixed}) over the joint pdf of $\beta_1$ and $\beta_2$. Note that, it is not necessary to simplify $\bar{Q}_F$ and $\bar{Q}_D$ in this case as well for the reasons discussed in section \ref{detection_probabilities} for Eq. (\ref{bar_PFi}) and Eq. (\ref{bar_PDi}).

The derivation for $Q_{F}$ and $Q_D$ can be extended to $M$ number of CSUs by selecting the threshold $\tau$ as $\tau=\frac{\lambda}{M}(\hat{\sigma}_{w_1}^p+\hat{\sigma}_{w_2}^p+\cdots+\hat{\sigma}_{w_M}^p)$. Following a procedure similar to two CSUs, ${Q}_{F}$ and ${Q}_D$ for $M$ CSUs with fixed values of $\beta_1,\beta_2,\cdots\beta_M$ can be obtained as
\begin{equation}
\label{Q_F_P}
{Q}_{F}=Q\left(\frac{M\lambda\prod\limits_{i=1}^{M}\beta_i^\frac{p}{2}-G_p\sum_{i=1}^M\prod_{j=1,j\neq i}^M\beta_j^{\frac{p}{2}}}{\sqrt{\frac{K_p}{N}}\sqrt{\sum_{i=1}^M\prod_{j=1,j\neq i}^M\beta_j^p}}\right),
\end{equation}
\begin{equation}
\label{Q_D_P}
{Q}_{D}=Q\left(\frac{M\lambda\prod\limits_{i=1}^{M}\beta_i^\frac{p}{2}-\sum\limits_{i=1}^M G_p\left(1+\beta_i\gamma_i\right)^\frac{p}{2}\prod\limits_{j=1, j\neq i}^M\beta_j^{\frac{p}{2}}}{\sqrt{\frac{K_p}{N}}\sqrt{\sum\limits_{i=1}^M\left(1+\beta_i\gamma_i\right)^p{\prod\limits_{j=1,j\neq i}^M}\beta_j^p}}\right),
\end{equation}
respectively. Finally, by averaging the $Q_{F}$ and $Q_D$ over the joint pdf of $\beta_1,\beta_2,\cdots,\beta_M$ we get $\bar{Q}_{F}$ and $\bar{Q}_D$ for general case of $M$ CSUs.

Once we obtain $\bar{Q}_F$ and $\bar{Q}_D$, we can derive the SNR wall. Once again, we derive the SNR wall for $M=2$ and then extend it to any $M$. 
Since deriving an expression for SNR wall is mathematically involved when we consider any $p$, we do it for $p=2$ only. However, we show that SNR wall is independent of $p$ using our simulation in section \ref{results}. 
Considering $p=2$ and using $\bar{Q}_F$ and $\bar{Q}_D$, to satisfy both the conditions in Eq. (\ref{condition}), the $\lambda$ has to be chosen such that
\begin{equation}
\small
\label{condition3}
\frac{G_p\left(10^{\frac{L_1}{10}}+10^{\frac{L_2}{10}}\right)}{2}\leq \lambda \leq \frac{G_p\left(\gamma_1+\gamma_2+10^{\frac{-L_1}{10}}+10^{\frac{-L_2}{10}}\right)}{2},
\end{equation}
Using Eq. (\ref{condition3}), the condition on $\gamma_1$ and $\gamma_2$ for $M=2$ can be obtained as
\begin{equation}
\label{condition4}	
\gamma_1+\gamma_2 \geq 10^{\frac{L_1}{10}}+10^{\frac{L_2}{10}}-10^{\frac{-L_1}{10}}-10^{\frac{-L_2}{10}}.
\end{equation}
Taking the equality sign in Eq. (\ref{condition4}) gives the the SNR wall.
Looking at the Eq. (\ref{condition4}), one can arrive at the following conclusions.
\begin{itemize}
	\item In the absence of NU at both the CSUs, i.e., $L_1=L_2=0$, one can always find a threshold for which both the conditions in Eq. (\ref{condition}) are satisfied at any SNR greater than zero, i.e., $\gamma_1,\gamma_2\geq 0$. This means that the sensing scheme is unlimitedly reliable when there is no NU. A similar result is shown in \cite{sanket_apcc} where they do not consider cooperation.
	\item When both the CSUs have the same NU, i.e., $L_1=L_2$ and the same SNRs, i.e., $\gamma_1=\gamma_2=\gamma$, then there is no improvement in terms of SNR wall when compared to using no CSS. For example, SNR wall as given in \cite{sanket_apcc}, i.e., $\gamma \geq 10^{\frac{L}{10}}-10^{\frac{-L}{10}}$ takes a value of $0.4646$ when $\beta$ (in dB) is in the range $[-1 \ 1]$, indicating that we require SNR of at least $0.4646$ to obtain unlimitedly reliable sensing. Now if we consider CSS with two CSUs having the same noise uncertainties and SNRs, we still require an SNR of $0.4646$ at both the CSUs in order to satisfy the SNR wall condition given in Eq. (\ref{condition}), indicating no improvement.
	\item The advantage of using CSS lies in the fact that the SNR wall is determined by combined SNR, i.e., $\gamma_1+\gamma_2$. Hence, even if the SNR is low at one CSU, we could still satisfy condition in Eq. (\ref{condition4}) by having sufficiently high SNR at other CSU and achieve unlimitedly reliable sensing. For example with $L_1=1 \ dB$ and $L_2=0.5 \ dB$, the combined SNR required for unlimited reliability is $0.6954$. This can be satisfied if one of the CSUs has the SNR of $0.3954$ and the other has a low SNR of $0.3$.
\end{itemize}
Using Eq. (\ref{Q_F_P}) and Eq. (\ref{Q_D_P}), one can obtain the SNR wall for $M$ CSUs as
\begin{equation}
\label{SNR_wall_P}
\sum_{i=1}^M\gamma_i=\sum_{i=1}^M 10^{\frac{L_i}{10}}-\sum_{i=1}^M 10^{\frac{-L_i}{10}},
\end{equation}
and the conclusions that are drawn for the case of $M=2$ can be easily extended to general case of $M$ CSUs.

We would like to mention here that we have not considered the fading in our system model mentioned in section \ref{system}. One may include the fading to make it more general. However this inclusion considerably increases the mathematical complexity since the expressions for $\bar{Q}_D$ will have multiple integrals in order to carry out the averaging over the pdfs of NU and instantaneous SNR under fading. For example, with $M=2$, the expression for $\bar{Q}_D$ consists of two integrals for hard combining and that for soft combining will have four integrals. Due to this, the approach that we have used in section \ref{SNR_wall} for deriving the SNR wall can no longer be used. It makes it hard and mathematically too involved to arrive at closed form of SNR wall expressions. One can still obtain the SNR wall in this case by numerical means by choosing the threshold for which the $\bar{Q}_F$ becomes $0$ which can be easily obtained using the expressions for $\bar{Q}_F$. This threshold can then be used to find the SNR for which the $\bar{Q}_D$ becomes $1$ which gives us the SNR wall. Due to the page limitation, we have not included the analysis and the simulation plots considering fading here.
\section{Results and Discussions}
\label{results}
\begin{figure}
	\centering
	\begin{subfigure}{.23\textwidth}
		\centering
		\includegraphics[width=4.6cm,height=5cm]{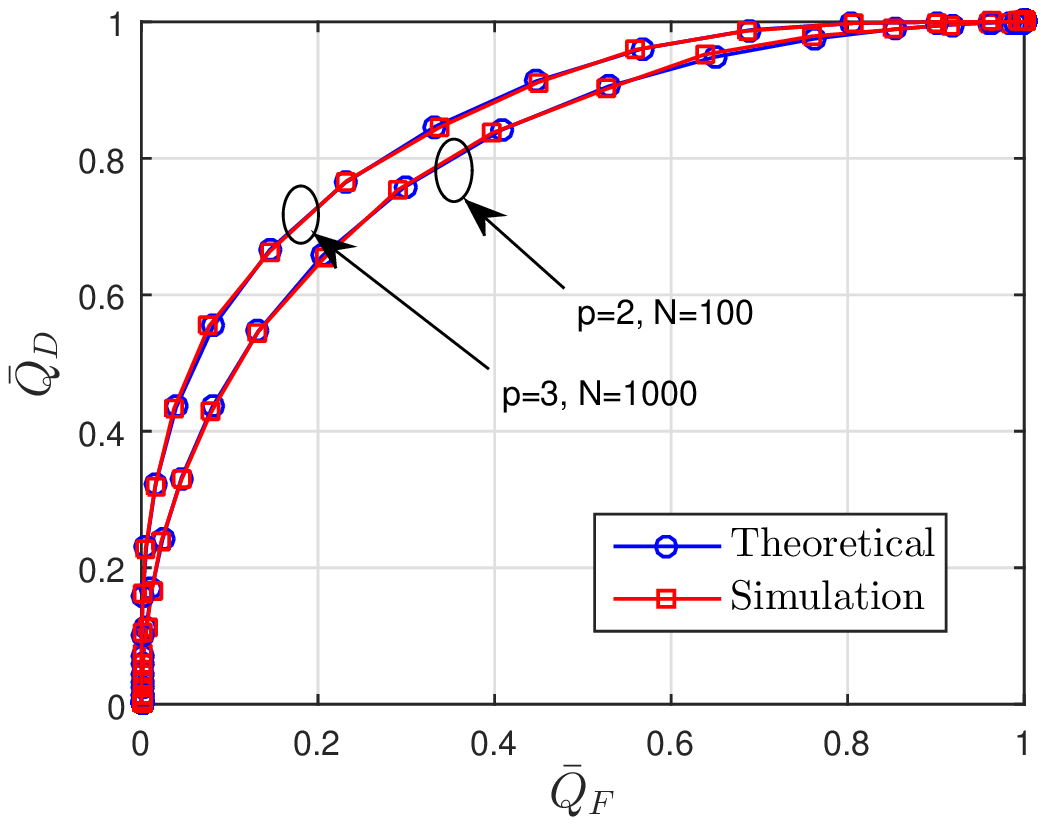}
		\caption{}
		\label{validation_k_out_M}
	\end{subfigure}%
	\begin{subfigure}{.23\textwidth}
		\centering
		\includegraphics[width=4.6cm,height=5cm]{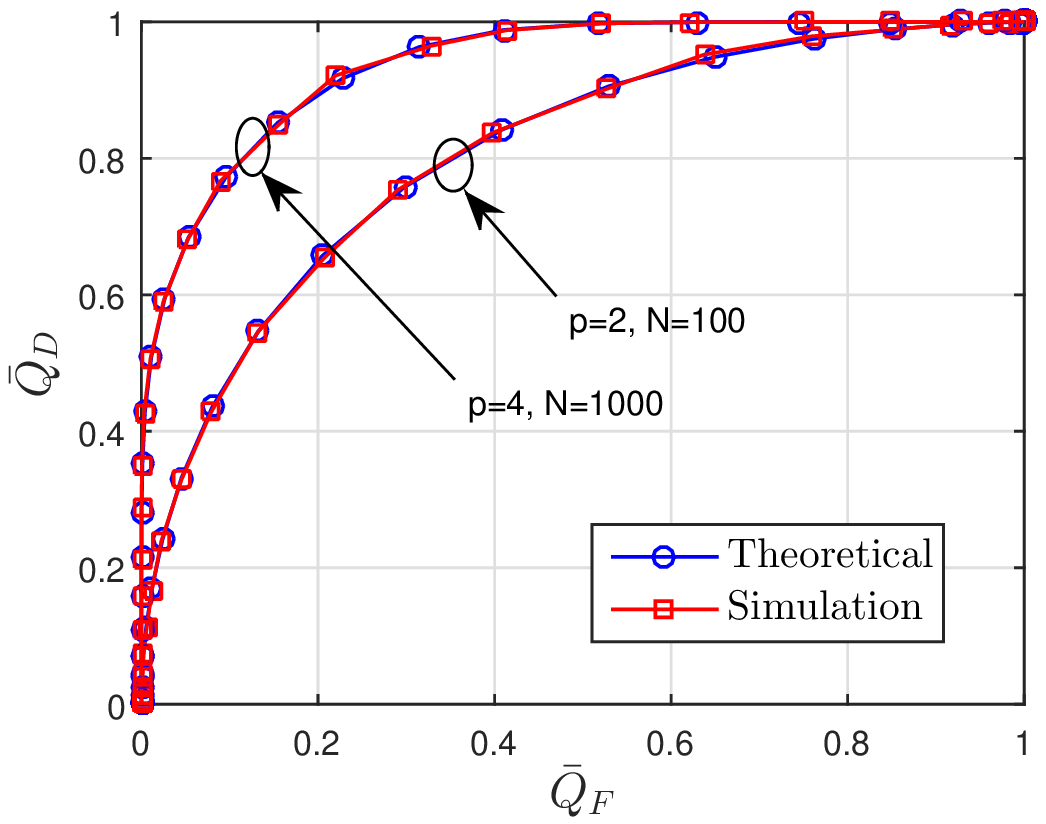}
		\caption{}
		\label{validation_soft}
	\end{subfigure}
	\caption{$\bar{Q}_F$ Vs. $\bar{Q}_D$ plots using Theoretical analysis and Monte Carlo simulations for (a) $k$ out of $M$ combining rule with $L_1=1 \ dB$, $L_2=0.7 \ dB$, $L_3=0.5 \ dB$, $\gamma_1=-5 \ dB$, $\gamma_2=-10 \ dB$, $\gamma_3=-15 \ dB$ and (b) soft combining with $L_1=1 \ dB$, $L_2=0.5 \ dB$, $\gamma_1=-5 \ dB$ and $\gamma_2=-15 \ dB$.}
	\label{validation}
\end{figure}
\begin{figure}[htb]
	\begin{minipage}[b]{1.0\linewidth}
		\centering
		\centerline{\includegraphics[width=8cm, height=5cm]{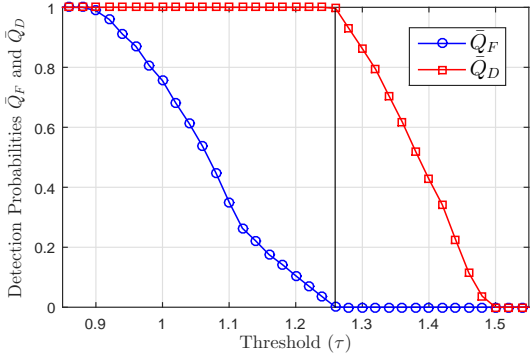}}
		\caption{Plots of $\tau$ Vs. $\bar{Q}_{F} \ \text{and} \ \bar{Q}_D$ for OR combining rule. Here, $N=10^6$, $M=2$, $L_1=1 \ dB$, $L_2=0.5 \ dB$, $p=2$, $\gamma_1=0.2$ and $\gamma_2=0.3676$.}
		\label{OR_wall}
	\end{minipage}
\end{figure}
\begin{figure}[htb]
	\begin{minipage}[b]{1.0\linewidth}
		\centering
		\centerline{\includegraphics[width=8cm, height=5cm]{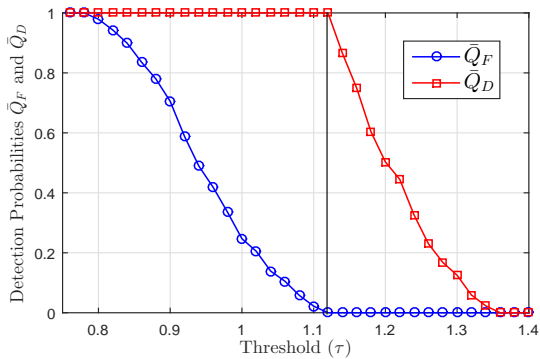}}
		\caption{Plots of $\tau$ Vs. $\bar{Q}_{F} \ \text{and} \ \bar{Q}_D$ for AND combining rule. Here, $N=10^6$, $M=2$, $L_1=1 \ dB$, $L_2=0.5 \ dB$, $p=2$, $\gamma_1=0.3277$ and $\gamma_2=0.2308$.}
		\label{AND_wall}
	\end{minipage}
\end{figure}
\begin{figure}[htb]
	\begin{minipage}[b]{1.0\linewidth}
		\centering
		\centerline{\includegraphics[width=8cm, height=5cm]{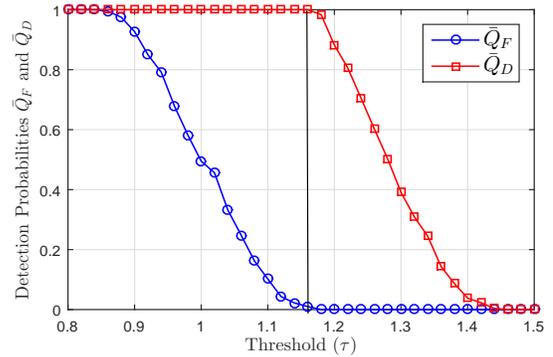}}
		\caption{Plots of $\tau$ Vs. $\bar{Q}_{F} \ \text{and} \ \bar{Q}_D$ for $k$ out of $M$ combining rule. Here, $N=10^6$, $M=3$, $L_1=1 \ dB$, $L_2=0.7 \ dB$, $L_3=0.5 \ dB$, $p=2$, $\gamma_1=0.2$, $\gamma_2=0.3238$, and $\gamma_3=0.2836$.}
		\label{k_out_M_wall}
	\end{minipage}
\end{figure}

In this section, first we validate the expressions for $\bar{Q}_F$ and $\bar{Q}_D$ for both hard and soft combining using the receiver operating characteristic (ROC) plots, i.e., $\bar{Q}_F$ Vs. $\bar{Q}_D$, obtained using expressions and Monte Carlo (MC) simulations. We then verify the analytical expressions for SNR wall derived in section \ref{SNR_wall} using MC simulations. Verification is done using the combined plots of threshold $\tau$ Vs. $\bar{Q}_F$ and $\bar{Q}_D$ by considering a very large value of $N=10^6$. As discussed in section III, the SNR wall for hard decision combining is independent of $p$ and hence all our plots on hard combining (Fig. \ref{OR_wall} to \ref{k_out_M_wall}) are shown for $p=2$ only. For MC simulation we generate PU signal as complex Gaussian with mean $0$ and variance $\sigma_s^2$. The noise at all the CSUs are generated as complex Gaussian with mean $0$ and variance $\sigma_{w_i}^2$, where $i=1,2,\cdots,M$. The results are averaged over $10^5$ realizations.  
In each iteration, noise samples are generated with variance $\sigma_{w_i}^2$ where $\sigma_{w_i}^2=\frac{\hat{\sigma}_{w_i}^2}{\beta_i}$, with $\hat{\sigma}_{w_i}^2=1$ and by taking the samples of $\beta_i$ from pdf given in Eq. (\ref{pdf_B_i}).
Note that, since we assume here $\hat{\sigma}_{w_1}^2=\hat{\sigma}_{w_2}^2=\cdots,\hat{\sigma}_{w_M}^2=1$, we get $\tau=\lambda$ and hence the plots of $\tau$ Vs. $\bar{Q}_F$ and $\bar{Q}_D$ are shown for verification of derived results. If we assume different $\hat{\sigma}_{w_i}^2$ at different CSUs, it is more appropriate to plot $\lambda$ Vs. $\bar{Q}_F$ and $\bar{Q}_D$ since in this scenario $\tau$ is different for all the CSUs but the $\lambda$ remains the same. 

In Fig. \ref{validation_k_out_M}, we show the ROC plots for $k$ out of $M$ combining rule using both theoretical analysis and MC simulations by considering $p=2$ and $p=3$. To obtain these plots using theoretical analysis we use Eq. (\ref{Q_F_k_out_M}) and Eq. (\ref{Q_D_k_out_M}). Overlapping of the plots conforms the correctness of our analysis. Since OR and AND combining rules are special cases of $k$ out of $M$ combining rule, we avoid showing the plots for them. Similar plots are shown for soft combining in Fig. \ref{validation_soft} where the expressions for $\bar{Q}_F$ and $\bar{Q}_D$ in section \ref{soft_derivation} for $M=2$ are used. Once again, the overlapping of the plots obtained using both the methods validates the theoretical analysis.

We next show the plots of $\tau$ Vs. $\bar{Q}_F$ and $\bar{Q}_D$. We start with one of the hard decision combining, i.e., OR combining rule, as discussed in section \ref{OR_rule} and consider two CSUs having $L_1=1 \ dB$ and $L_2=0.5 \ dB$. Substitution in Eq. (\ref{OR_SNR_Wall}) leads to the SNR wall as $\gamma_1=0.4646$ or $\gamma_2=0.3676$. In Fig. 1, we show the plot of $\tau$ Vs. $\bar{Q}_F$ and $\bar{Q}_D$ by considering $\gamma_1=0.2$ which is less than the SNR wall of $0.4646$ and $\gamma_2=0.3676$. This corresponds to choosing one of the SNRs as less than the corresponding SNR wall and the other one as the value satisfying SNR wall condition. The vertical line in Fig. 1 gives the threshold for which both the conditions in Eq. (\ref{condition}) are satisfied. This means, if we set the threshold as $1.26$ and choose a large value of $N$, we can achieve $\bar{Q}_F=0$ and $\bar{Q}_D=1$, i.e., unlimited reliability. This indicates that if $\gamma_1\geq0.4646$ or $\gamma_2\geq0.3676$, it is possible to find a threshold to achieve unlimited reliability.     

The condition for SNR wall in the case of AND combining rule is obtained using Eq. (\ref{SNR_wall_AND}). Using the same $L_1$ and $L_2$ as in the OR case, we obtain $\gamma_1=0.3277$ and $\gamma_2=0.2308$. In this case, both these SNRs must be $\geq$ these values in order to achieve the unlimited reliability. The plot for AND combining rule is shown in Fig 2 using the SNR values equal to their SNR walls. The vertical line in the plot shows that using $\tau=1.12$ one can achieve unlimited performance when $N$ is very large.

In Fig. 3, we demonstrate the SNR wall for $k$ out of $M$ combining rule and consider three CSUs, i.e., $M=3$, having $L_1=1 \ dB$, $L_2=0.7 \ dB$ and $L_3=0.5 \ dB$ with $k=2$. Using these parameters in Eq. (\ref{SNR_Wall_K_out_M}), we compute the SNR walls as $\gamma_1=0.3806$, $\gamma_2=0.3238$ and $\gamma_3=0.2836$. In Fig. 3, we show the plots by choosing $\gamma_1=0.2$ which is below the required value of SNR wall and $\gamma_2=0.3238$ and $\gamma_3=0.2836$ which are equal to their SNR walls. Since we have $k=2$, and $2$ out of $3$ CSUs have the inputs with SNR $\geq$ their SNR walls, an unlimited operation is obtained. We can see from Fig. 3 that choosing a value of $\tau=1.16$ (threshold corresponding to the vertical line) gives the unlimited reliability, i.e., choosing this threshold value with two of the three SNRs $\geq$ their SNR walls gives us $\bar{Q}_F=0$ and $\bar{Q}_D=1$. 

\begin{figure}[htb]
	\centering
	\includegraphics[width=8cm, height=5cm]{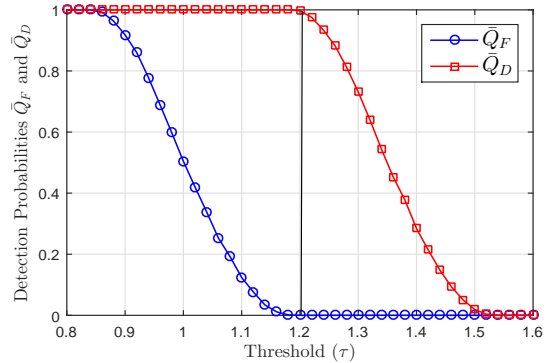}
	\caption{Plots of $\tau$ Vs. $\bar{Q}_{F} \ \text{and} \ \bar{Q}_D$ for soft combining. Here, $N=10^6$, $M=2$, $L_1=1 \ dB$, $L_2=0.5 \ dB$, $p=2$, $\gamma_1=0.3$ and $\gamma_2=0.3954$.}
	\label{soft_wall}
\end{figure}

As done for the hard decision combining, in Fig. 4, we show plots to validate the SNR wall expressions when soft decision combining is used. Once again we consider $M=2$ and same uncertainties as used in OR combining rule. The conditions on $\gamma_1$ and $\gamma_2$ can be obtained by substituting these values in Eq. (\ref{condition4}) which gives us SNR wall as $\gamma_1+\gamma_2=0.6954$. It shows that in order to get the unlimited reliability, we require the combined SNR at two CSUs to be $\geq 0.6954$. In Fig. 4, we choose $\gamma_1=0.3$ and $\gamma_2=0.3954$, thus satisfying the SNR wall condition. We see that setting $\tau=1.2$ gives us the unlimited performance which is demonstrated by a vertical line at $\tau=1.2$. This shows that even if the SNR is low at one SU and the other has sufficiently high SNR, we still get the unlimitedly reliable sensing. 
\begin{figure}[htb]
	\centering
	\includegraphics[width=8cm, height=5cm]{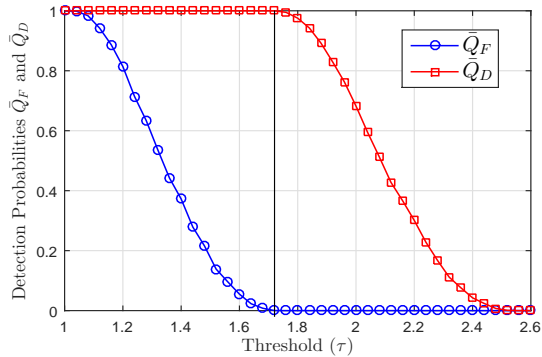}
	\caption{Plot of $\tau$ Vs. $\bar{Q}_{F} \ \text{and} \ \bar{Q}_D$ for soft combining. Here, $N=10^6$, $M=2$, $L_1=1 \ dB$, $L_2=0.5 \ dB$, $p=3$, $\gamma_1=0.3$ and $\gamma_2=0.3954$.}
	\label{soft_wall_p_3}
\end{figure}

In Fig. 5, we demonstrate that the SNR wall is independent of the value of $p$ when we use soft combining. Here, we choose $p=3$ instead of $p=2$ and the other parameters are kept the same as in Fig. 4. The vertical line at $\tau=1.72$ shows that setting the threshold at $1.72$ and taking a large $N$ gives us the unlimited performance. This clarifies that the SNR wall is independent of the value of $p$, since we get unlimited performance by using the same SNR values as used when $p=2$ in Fig. 4. 
We would like to mention here that due to the page limitations, we have included only the interesting results for each case. Other special cases can also be validated using the similar procedure.  

\section{Conclusion}
In this work, we study cooperative spectrum sensing when all the CSUs employ generalized energy detector under noise uncertainty. We derive the SNR wall for hard as well as soft decision combining. For hard combining, we consider all three possible cases, i.e., OR, AND and $k$ out of $M$. 
For soft combining, we consider equal gain combining and derive the SNR wall for the same. We also validate all our theoretical analysis with Monte Carlo simulations. Our future research work involves analysis of SNR wall for cooperative spectrum sensing when the cooperating secondary users experience fading.






%



\bibliographystyle{IEEEtran}
\bibliography{mybib}

\end{document}